# Generalized Einstein relation for disordered semiconductors – implications for device performance

Yohai Roichman and Nir Tessler[a]

Electrical Engineering Dept., Technion Israel institute of technology, Haifa 32000, Israel.

**Abstract**

The ratio between mobility and diffusion parameters is derived for a Gaussian-like density of states. This steady-state analysis is expected to be applicable to a wide range of organic materials (polymers or small molecules) as it relies on the existence of quasi-equilibrium only. Our analysis shows that there is an inherent dependence of the transport in trap-free disordered organic-materials on the charge density. The implications for the contact phenomena and exciton generation rate in light emitting diodes as well as channel-width in field-effect transistors is discussed.

[a] Email: nir@ee.technion.ac.il. W.pg: http://tiger.technion.ac.il/~nir/

It is well established that charge transport in disordered semiconductors is not fully characterized by the conventional current continuity equations [1]. In the context of organic semiconductors it is accepted that the mobility is exponential in the square root of the electric field [2] however this modification is also often found insufficient. The deviation of the experimental results from the conventional current continuity equations description is typically explained either in the context of transport under non-equilibrium conditions [3,4] or with the aid of detailed Monte Carlo simulations [5] of hopping transport. A common feature of these detailed studies in that they do not consider (or even preclude) effects associated with the charge density. One of the quoted indications for the standard continuity equations being incomplete is the finding that the ratio between the diffusion coefficient of charge carriers and the mobility is larger [6,7] than the classical Einstein relation of kT/q or the presence of "fast-carriers"[8]. In this work we suggest a basic explanation, which relies on two, commonly used, assumptions only. a) The density of states (DOS) can be described using a Gaussian. b) The charge carriers can be described as being at equilibrium conditions (i.e. one can use Fermi-Dirac statistics).

The relation between the diffusion coefficient (D) and the mobility (μ) of in the low density limit is given by D/μ=kT/q [9](Einstein relation) where k is Boltzmann coefficient, q is the charge of the particle and T is the characteristic temperature. A generalized relation between the diffusion coefficient and the mobility (i.e. generalized Einstein relation) can be derived for a general charge-carrier energy-distribution function, and a general DOS function [9]:

1) $$\frac{D}{\mu} = \frac{p}{q\frac{\partial p}{\partial \eta}}$$

Where p is the particle concentration and η is the chemical potential.

The general Einstein relation has so far been derived for high-density (degenerate) crystalline-semiconductors[9] or degenerate-semiconductors having a band tail [10], i.e. a distribution of states in the forbidden gap. In this paper the generalized Einstein relation is calculated for a Gaussian density of states (NOT a tail), as often assumed in the case of organic semiconductors and few other amorphous semiconductors. The functional form of a Gaussian DOS is given by equation 2:

$$2)\quad DOS(\varepsilon) = \frac{N_V}{\sqrt{2\pi} \cdot a} \exp\left[-\left(\frac{\varepsilon - \varepsilon_0}{\sqrt{2} \cdot a}\right)^2\right]$$

Where $\varepsilon$ is the normalized energy $\varepsilon = \frac{E}{kT}$, $\varepsilon_0$ is the Gaussian center, $N_V$ is the effective DOS, and $a$ is defined as the normalized Gaussian variance: $a \equiv \frac{\sigma_0}{kT}$ ($\sigma_0$ is the variance of the energy Gaussian distribution). If we Let $f(\varepsilon, \xi)$ be the normalized Fermi-Dirac distribution function then the charge concentration can be written as:

$$3)\quad p(\xi) = \int_{-\infty}^{\infty} DOS(\varepsilon) f(\varepsilon, \xi) d\varepsilon$$

Inserting 2 and 3 into 1, one can derive an implicit relation between the diffusion coefficient and the mobility (generalized Einstein relation) for the case of a Gaussian DOS:

$$4)\quad \frac{D}{\mu} = \frac{kT}{q} \frac{\int_{-\infty}^{\infty} \exp\left[-\left(\frac{\varepsilon - \varepsilon_0}{\sqrt{2} \cdot a}\right)^2\right] \cdot \frac{1}{1 + \exp(\varepsilon - \xi)} d\varepsilon}{\int_{-\infty}^{\infty} \exp\left[-\left(\frac{\varepsilon - \varepsilon_0}{\sqrt{2} \cdot a}\right)^2\right] \cdot \frac{\exp(\varepsilon - \xi)}{[1 + \exp(\varepsilon - \xi)]^2} d\varepsilon}$$

It can be shown that when the major part of the charge energy-distribution is far from the chemical potential (i.e. the Fermi-Dirac distribution can be approximated by a Boltzmann distribution) the generalized Einstein relation approaches the classical value ($D/\mu=kT/q$). This non-degeneracy condition is applicable when the Gaussian variance ($a$) is small and the chemical potential is distant more than 5kT from the Gaussian center (Figure 1). A more realistic Gaussian variance to describe organic semiconductors is in the order of 70-150 meV ($a$=3-6 at room temperature). In such a case the charge concentration peak remains near the chemical potential even when it is 20kT below the center of the Gaussian, $E_0$. This implies that an amorphous organic semiconductor is always degenerate down to very low charge concentration and the classical Einstein relation is not expected to hold under any realistic condition. Namely, the generalized Einstein relation has to be calculated in its full form.

Calculation of the Einstein relation versus the position of the normalized chemical potential ($\xi$) is given for different DOS variances ($a$) in Figure 2 (Notice that the inverse of the Einstein relation [μ/D] is shown). As expected, for very low chemical potential level (very low charge density) the generalized Einstein relation tends towards its classical value of 1 (kT/q). In common devices however, the chemical potential is expected to be relatively high. In this case, as the DOS width (disorder) increases, the Einstein relation increases and the diffusion coefficient becomes considerably larger than kT/q*μ. Also, when the chemical potential goes up (for a given width, *a*), meaning the charge concentration increases, the Einstein relation increases as well. The dependence of the Einstein relation on the relative charge density ($p/N_V$) can be seen clearly in Figure 3. Namely, materials that are loosely packed are more likely to exhibit a large Einstein relation values.

Next, we examine the impact on device performance and/or analysis. The most obvious device configuration employing a high charge density is that of the field-effect transistor (FET) that operate at charge density of ~$10^{18}$-$10^{19}$cm$^{-3}$ ($p/N_V$=0.01-0.1). Taking the effects discussed here the channel width would be well beyond the monolayer deduced based on the classical Einstein-relation[11]. Namely, in the analysis of charge transport in FET one should consider more then the interface monolayer and when chain-alignment is attempted one must align more then just the interface layer. Similar charge densities exist at the contact interface of space charge limited light emitting diodes (LEDs). Since the effect of the space charge is to make the current near the contact mainly diffusive the effect of the generalized Einstein-relation is of primary importance to the understanding of the contact phenomena and its functional dependence on various parameters (temperature etc.) and will be described elsewhere. The effect on the bulk of an LED can be estimated assuming DOS variance, charge concentration, and total state concentration of ~100 meV (*a*=4), $p \approx 10^{16}$cm$^{-3}$, and $N_V \approx 10^{20}$-$10^{21}$cm$^{-3}$ respectively. Under these conditions the Einstein relation is about twice its classical value of kT/q, and the diffusion coefficient will be larger by the same ratio.

To test whether this mechanism can help to explain the too broad rise of the measured LED turn-on presented by Pinner et. al. (figure 15 in [7]) we reproduced the simulations and accounted for the effect of the DOS being Gaussian like. Figure 4a shows the turn on dynamics of a 60nm long LED where the assumed mobility values are $1 \times 10^{-4}$cm$^2$V$^{-1}$s$^{-1}$ and $1 \times 10^{-5}$cm$^2$V$^{-1}$s$^{-1}$ for holes and electrons respectively. The net applied voltage ($V_{Appl}-V_{bi}$) is assumed to be 4V. All other parameters are as in [7]. The bottom line was calculated for a negligibly narrow DOS, where the classical Einstein relation holds, and the upper curve was calculated for a DOS width of 7kT. If we extract a turn on time ($t_d$) from Figure 4a we find td=48ns and td=42ns for the narrow and wide DOS cases, respectively. Using the arrival time of the front-end of the distribution (light turn-on) one may deduce hole mobility of $1.9 \times 10^{-4} cm^2 V^{-1} s^{-1}$ and $2.2 \times 10^{-4} cm^2 V^{-1} s^{-1}$ for the narrow and wide DOS cases, respectively. If we apply the method outlined in [7] to extract the time where the body of the hole distribution has arrived at the cathode (see Figure 4b) we find that for both cases the arrival time is about 72ns and the deduced mobility is $\mu = \dfrac{d^2}{V \Delta t} = 1.25$ $\times 10^{-4} cm^2 V^{-1} s^{-1}$ which is very close to the mobility used for both simulations (the extra 25% is due to the space charge enhanced electric field [7] in the bulk that is not accounted for in such a simple expression). Namely, when the Gaussian nature of the DOS is taken into account the charge-carrier front is broadened giving rise to a longer rise in the curve. These differences are consistent with those found between "classical" simulations and experiment in [7] and hence we conclude that the modified Einstein relation agrees well with experimental findings. In other words, it strengthens the notion that there is no need to resort to non-equilibrium effects in order to describe (or extract parameters from) the organic devices considered here. Moreover, this calculation demonstrates the method of mobility extraction developed in [7] as being largely immune to "fast-carriers" effects. The self-consistent nature of the numerical model also reveals another feature of the DOS induced diffusion enhancement. As the charge recombination is Langevin [7,12] (diffusion controlled) the enhanced diffusion-rate enhances the exciton generation-rate and hence the light-output is enhanced as well (see Figure 4a).

Finally, we chose the Gaussian shape for the DOS-function as it is likely to be a good approximation for reasonably moderate charge density and has been shown to successfully describe experimental findings [13]. At low charge concentration the description might not be so adequate and a modified DOS should probably be taken into account, where there is a cutoff of states. Such an approach was discussed in the context of the transition, in time, from dispersive to non-dispersive transport [4]. Part of the justification/motivation here is related to the scale of the device. For example in a 1μm thick device a density of $10^{12}$ $cm^{-3}$ corresponds to a single state in the device and hence there would be an effective cutoff at this density. To illustrate this effect we numerically computed in Figure 2 and Figure 3 the Einstein relation for a Gaussian DOS having a "cutoff" at either 40kT or 19kT below its center (~1eV or ~0.5eV at room temperature), dashed and dotted lines respectively. We note that this cutoff forces the material into a non-degenerate state for chemical potential below the cutoff energy.

To conclude, we have computed the effect of a disordered density of states on the diffusion-mobility ratio. The method is general [9] and relies on the shape of the DOS only. The one assumption regarding the dynamics of the system is that it is close to quasi-equilibrium so that one can define a chemical potential (as is in most device models). The good agreement with experiments strengthens the notion of quasi-equilibrium even for thin, 100nm wide, LEDs. Moreover, the diffusion-mobility ratios found are in good agreement with those deduced using detailed Monte Carlo simulations[6] both as a function of the DOS width ($a$) and as a function of temperature. We have also shown that this effect plays an important role for both LEDs and FETs. The results show that the charge density affects the transport phenomena in trap-free disordered semiconductors making the system nonlinear (with N).

*Acknowledgments*:

We acknowledge fruitful discussions with Harvey Scher. This research (No. 56/00-11.6) was supported by THE ISRAEL SCIENCE FOUNDATION. N.T. thanks the Israeli Board of Higher Education for an Allon fellowship.

Y. Roichman and N. Tessler 7# REFERENCES

[1] H. Scher, M. F. Shlesinger, and J. T. Bendler, Physics Today **44,** 26-34 (1991); M. Van der Auweraer, F. C. Deschryver, P. M. Borsenberger, and H. Bassler, Advanced Materials **6,** 199-213 (1994); Q. Gu, E. A. Schiff, S. Grebner, F. Wang, and R. Schwarz, Phys. Rev. Lett. **76,** 3196-3199 (1996).

[2] W. D. Gill, J. Appl. Phys. **43,** 5033 (1972).

[3] H. Scher and E. M. Montroll, Phys. Rev. B **12,** 2455–2477 (1975).

[4] E. M. Horsche, D. Haarer, and H. Scher, Phys. Rev. B **35,** 1273-1280 (1987).

[5] H. Bassler, G. Schonherr, M. Abkowitz, and D. M. Pai, Phys. Rev. B - Cond. Matt. **26,** 3105-3113 (1982).

[6] R. Richert, L. Pautmeier, and H. Bassler, Phys. Rev. Lett. **63,** 547-550 (1989).

[7] D. J. Pinner, R. H. Friend, and N. Tessler, J. Appl. Phys. **86,** 5116-5130 (1999).

[8] P. W. M. Blom and M. Vissenberg, Physical Review Letters **80,** 3819-3822 (1998).

[9] N. W. Ashcroft and N. D. Mermin, *Solid State Physics* (HOLT, RINEHART AND WINSTON, New York, 1988).

[10] D. Ritter, E. Zeldov, and K. Weiser, Phys. Rev. B. **38,** 8296-8304 (1988); K. P. Ghatak, A. K. Chowdhury, S. Ghosh, and A. A. N. Chakrav, Phys. Stat. Solid. (b) **99,** K55-K61 (1980).

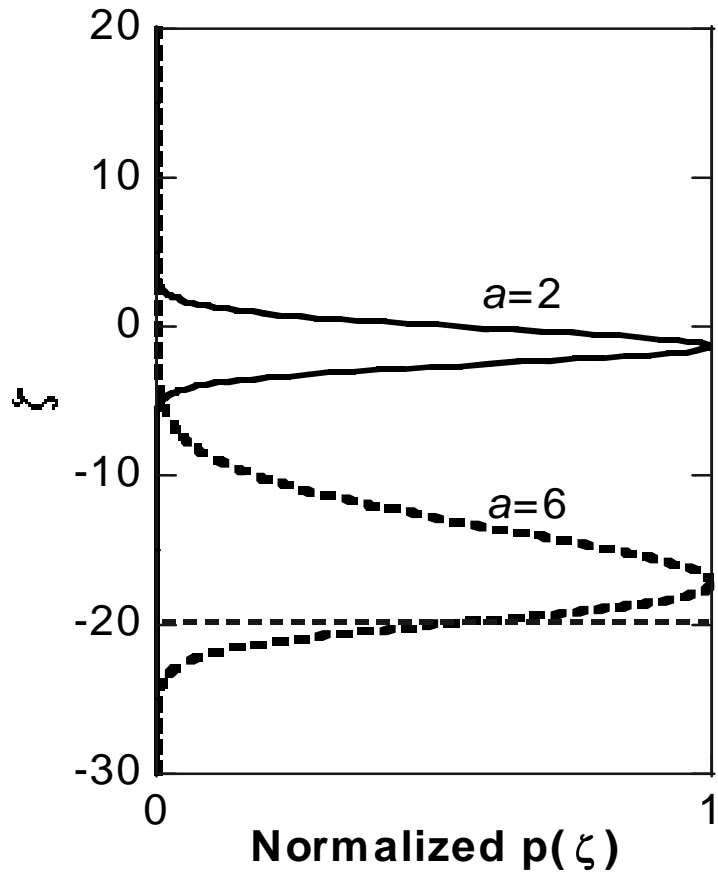

**Figure 1.** The charge concentration distribution in energy as calculated for the normalized chemical potential ($\eta$) being at $(E_0-20kT)/kT$. Top curve is calculated for gaussian variance of $a=2$ and bottom curve for $a=6$. For the low DOS variance ($a=2$) the charge concentration peak is close to the center of the DOS and hence is far from the chemical potential (non-degenerate case). On the other hand, for the higher DOS variance ($a=6$) the charge peak is adjacent to the chemical potential (the semiconductor is degenerate)

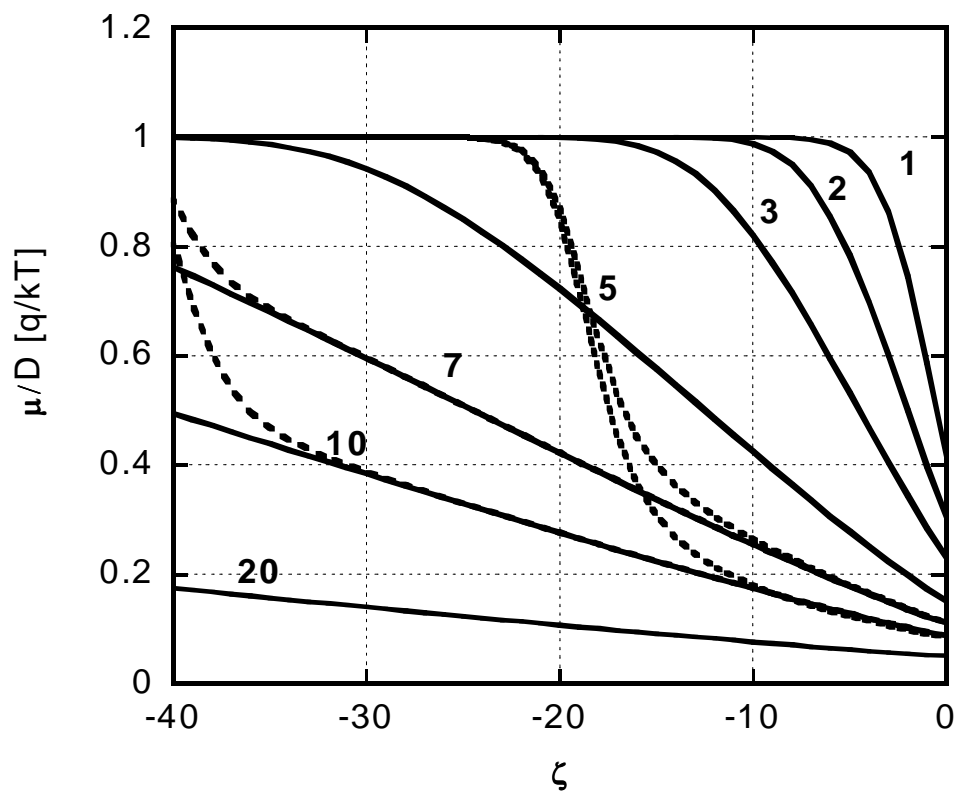

**Figure 2.** The inverse of Einstein relation (i.e. μ/D) versus the chemical potential (η) for different DOS variances (*a*). The solid lines were calculated for a Gaussian DOS. The dotted lines were calculated for a Gaussian DOS with cutoff at -19kT (~0.5eV at RT) and the dashed lines for a cutoff at -40kT (~1eV at RT). The cutoff effect is shown only for DOS width of *a*=7,10.

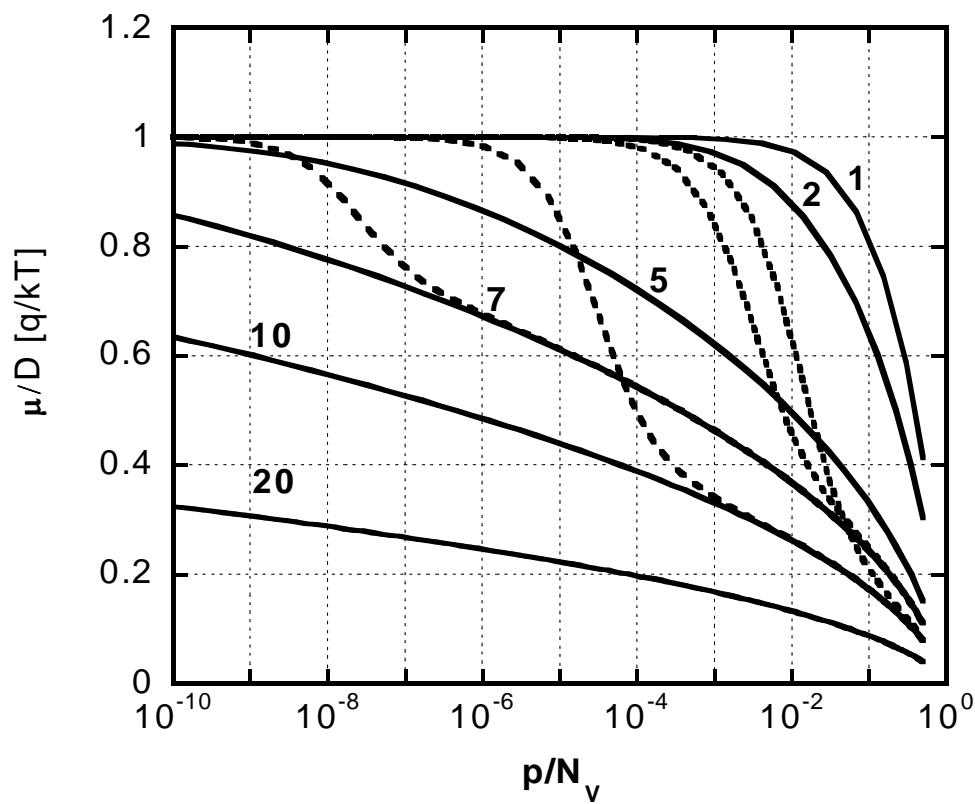

**Figure 3.** The inverse of Einstein relation (i.e. µ/D) versus the normalized charge concentration (p/$N_V$) for different DOS variances (*a*). The solid lines were calculated for a Gaussian DOS. The dotted lines were calculated for the Gaussian DOS with cutoff at –19kT (~0.5eV at RT) and the dashed lines for a cutoff at –40kT (~1eV at RT). The cutoff effect is shown only for DOS width of *a*=7,10.

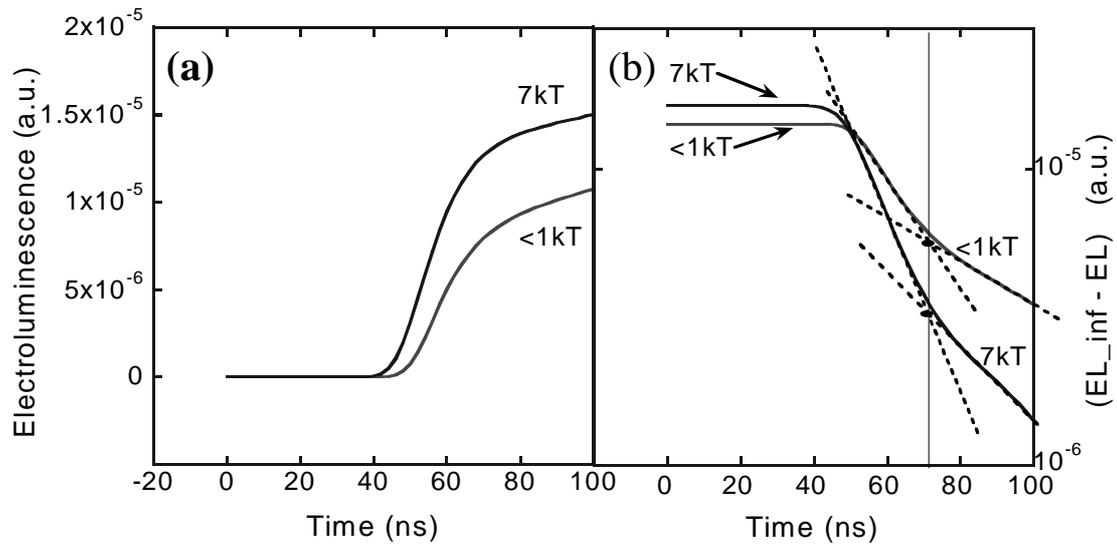

**Figure 4.** (a) Simulated light emission as a response to a step voltage pulse of 4V applied at t=0. The bottom curve was simulated for a material where the classical Einstein relation holds (as for σ<1kT and the top curve for a material having a Gaussian DOS with a width of σ=7kT at room temperature. Note the initial fast rise (hole arriving at the cathode) followed by a longer rise (electrons penetrate the device). (b) The data as in (a) but the curves values were extracted from the steady-state value and presented on a log scale (see Ref. 7). This way the transition between hole and electron dominated response is clearly visible and is marked by the vertical line.